# Combining hybrid and opaque scintillator techniques in the search for double beta plus decays

## A concept study for the NuDoubt$^{++}$ experiment

**Manuel Böhles**[1], **Sebastian Böser**[1], **Magdalena Eisenhuth**[1], **Cloé Girard-Carillo**[1], **Kitzia M. Hernandez Curiel**[1], **Bastian Keßler**[1], **Kyra Mossel**[1], **Veronika Palušová**[1], **Stefan Schoppmann**[2,a], **Alfons Weber**[1], **Michael Wurm**[1]

[1] Johannes Gutenberg-Universität Mainz, Institut für Physik, 55128 Mainz, Germany
[2] Johannes Gutenberg-Universität Mainz, Detektorlabor, Exzellenzcluster PRISMA$^+$, 55128 Mainz, Germany

**Abstract** Double beta plus decay is a rare nuclear disintegration process. Difficulties in its measurement arise from suppressed decay probabilities, experimentally challenging decay signatures and low natural abundances of suitable candidate nuclei. In this article, we propose a new detector concept to overcome these challenges. It is based on the first-time combination of hybrid and opaque scintillation detector technology paired with novel light read-out techniques. This approach is particularly suitable for detecting positrons (beta plus) signatures. We expect to discover two-neutrino double beta plus decay modes within 1 tonne-week exposure and are able to probe neutrinoless double beta plus decays at several orders of magnitude improved significance compared to current experimental limits.

## 1 Introduction

Neutrino-less double beta ($0\nu\beta\beta$) decay searches represent the most sensitive experimental access way to explore the Dirac or Majorana nature of the neutrino [1,2]. The current generation of double beta ($\beta\beta$) experiments is able to achieve lower half-life limits for neutrino-less double beta minus ($0\nu 2\beta^-$) decaying isotopes on the order of $10^{26}$ years [3–6]. However, these half-life limits will have to be improved by several orders of magnitude to explore the effective Majorana neutrino mass scale associated with the normal neutrino mass ordering.

This evolution in sensitivity will require experiments based on considerably larger $\beta\beta$ isotope masses (tens of tonnes), excellent energy resolution to suppress the tail of the standard model double beta ($2\nu\beta\beta$) decay spectrum and excellent radiopurity or background discrimination to maintain a clean region of interest (ROI) around the Q-value [2]. In this paper, we present the novel NuDoubt$^{++}$ (**Neu**trino **Dou**ble **bet**a **plus plus**) detector concept with the potential to fulfill all three of these demands. In continuation of the very successful development path taken by the KamLAND-ZEN and SNO+ experiments [7,8], we explore the advancement of isotope-loaded liquid scintillator detectors to higher loading factors, better energy resolution and improved particle ID capabilities. In particular, we explore the combination of two novel techniques for liquid scintillator detectors: hybrid scintillators that permit to measure the ratio of Cherenkov and scintillation signals and opaque scintillators that can provide centimetre-scale event topology information [9,10]. Combining both approaches in a single detector not only greatly enhances the background discrimination capabilities compared to regular organic scintillators, but also creates avenues for heavy isotope loading with minimal impact on the scintillation output and hence energy resolution of the experiment.

Double beta searches have been proposed individually for hybrid and opaque scintillators a few years ago, while the idea to exploit the unique signature of four or two annihilation gamma-rays for background suppression in search of double positron emission ($2\nu 2\beta^+$) or simultaneous electron capture and positron emission ($2\nu EC\beta^+$), respectively, dates back several decades [11]. Early ideas on hybrid detection published in 2016 involved the use of prompt directional Cherenkov light in addition to isotropic scintillation light to discriminate double electron decays from solar B-8 neutrino background [12,13]. In 2022, an approach using slow scintillators that improve separation of Cherenkov and scintillation light was published [14]. In the same year, our idea to exploit the ratio of Cherenkov and scintillation light in an hybrid detector for the search of double positron emission was disseminated [15]. Before, sensitivity towards $0\nu 2\beta^-$ in a kilotonne-scale hybrid detector was shown in a simulation study published by the Theia collaboration in 2020 [9]. First demonstrations of the hybrid detector concept through small-scale prototypes were published since 2017 [16,17]. A first tonne-scale hybrid detector for double electron decays was put forward by the NuDot collaboration in 2018 based on earlier small-scale prototypes [18]. Also since 2018, the LiquidO collaboration disseminated for the first time the concept of searches for double weak decays exploiting distinct event topologies in opaque media suitable for double electron and double positron searches [19–21]. This idea is refined by the LiquidO collaboration [22]. Demonstrations of the opaque detector concept through small-scale prototypes were published by the LiquidO collaboration since 2021 [10,23–26].

[a] e-mail: stefan.schoppmann@uni-mainz.de (corresponding author)
URL: https://nudoubt.uni-mainz.de

The present article discusses both the general properties of a new hybrid-opaque $\beta\beta$ experiment and a first outline for a tonne-scale (scintillator mass) demonstrator experiment. The corresponding detector features a fiducial volume of one cubic metre of hybrid-slow opaque scintillator densely instrumented with novel optimised wavelength-shifting fibres. We show that such a detector is well suited to perform first measurements of currently little explored two-neutrino double weak decays, such as $2\nu 2\beta^+$ and $2\nu\mathrm{EC}\beta^+$, and set limits on the corresponding neutrinoless $0\nu\mathrm{EC}\beta^+$ and $0\nu 2\beta^+$ half-lives. The isotopes in question are Kr-78, Xe-124, and Cd-106. Thus, the prototype would provide both a demonstration of the essential detector properties and additional data points to support theoretical modeling of $0\nu\beta\beta$ nuclear matrix elements. If successful, $2\nu\mathrm{EC}\beta^+$ and $2\nu 2\beta^+$ decays would be the rarest standard model (SM) decay processes ever observed.

This article is organised as follows: We first present the motivation to search for $2\nu\mathrm{EC}\beta^+$ and $2\nu 2\beta^+$ decays in Sect. 2. Next, we detail the detection concept of our new approach in Sect. 3. In Sect. 4, we discuss suitable isotope candidates for double beta plus searches. In Sect. 5, we give special attention to a key component of our detector concept, a novel type of light guides. We then discuss our expected backgrounds in Sect. 6, propose a basic design in Sect. 7, and evaluate our sensitivity in Sect. 8, before we conclude in Sect. 9.

## 2 Motivation of searches for double beta plus decays

Double beta ($\beta\beta$) decay is a rare nuclear disintegration which changes the nuclear charge number $Z$ by two units while keeping the nucleon number $A$ unchanged [27]. It is a second-order weak interaction process characterised by extremely long half-lives. Double beta decay can occur in different modes based on the types of particles emitted during the decay. The two main modes of interest of $\beta\beta$ processes are two-neutrino ($2\nu\beta\beta$) double beta decay, in which two (anti)neutrinos are also appearing in the final state, and neutrinoless ($0\nu\beta\beta$) double beta decay without the emission of any neutrinos:

$$2\nu 2\beta^- : \quad (A, Z) \rightarrow (A, Z+2) + 2e^- + 2\overline{\nu} \tag{1}$$

$$0\nu 2\beta^- : \quad (A, Z) \rightarrow (A, Z+2) + 2e^- \tag{2}$$

$2\nu\beta\beta$ and $0\nu\beta\beta$ can be distinguished experimentally by the amount of energy the charged leptons deposit in a detector. In the two-neutrino case, the released energy is shared between the charged leptons and the neutrinos, resulting in a wide energy distribution of the charged leptons. In the neutrinoless case, the charged leptons carry all the energy, resulting in a mono-energetic peak.

The processes where two neutrinos are emitted are predicted by the SM, and $2\nu 2\beta^-$ decay has been observed in several nuclei. The $0\nu 2\beta^-$ decay has not yet been experimentally observed. In most models, it violates total lepton number conservation by two units and its observation would imply Majorana nature of neutrinos [1,2,28,29].[1] Therefore, it is a unique tool to probe physics beyond the SM. Considerable experimental and theoretical efforts have been directed towards studying the double electron emission mode, $2\beta^-$, given its potential as the most promising mode for potential detection of $0\nu\beta\beta$, with theoretical investigations focusing on understanding the underlying decay mechanisms and nuclear structure models [1,2,31,32]. However, in recent years, interest has been renewed in the less explored double beta transitions that decrease the nuclear charge number: double positron decay, $2\beta^+$, positron emitting electron capture, $\mathrm{EC}\beta^+$, and double electron capture, 2EC:

$$2\nu 2\beta^+ : \quad (A, Z) \rightarrow (A, Z-2) + 2e^+ + 2\nu \tag{3}$$

$$2\nu\mathrm{EC}\beta^+ : \quad (A, Z) + e^- \rightarrow (A, Z-2) + e^+ + 2\nu \tag{4}$$

$$2\nu 2\mathrm{EC} : \quad (A, Z) + 2e^- \rightarrow (A, Z-2) + 2\nu \tag{5}$$

The limited exploration of these modes arises from the suppressed decay probabilities, less favourable decay Q-values, experimental challenges in the observation of a signature of these decay modes and low natural abundances of suitable candidate nuclei [33–37]. The identification of 2EC can be achieved by detecting the cascade of X-rays and Auger electrons generated during the filling of vacancies subsequent to the capture of two atomic electrons. The nuclear binding energy, Q, released during this process (typically around 1 MeV) is predominantly carried away by the two neutrinos, which remain undetected within the detector. For decay modes associated with positrons, the experimental signature is the simultaneous emission of one (in case of EC$\beta^+$) or two (in case of 2$\beta^+$) pairs of 511 keV annihilation gamma-rays.

[1] An alternative model for neutrinoless double beta decay without the existence of Majorana neutrinos or lepton number violation has been proposed e.g. in [30].

Although all three of the nuclear charge decreasing two-neutrino double beta modes (reactions 3, 4 and 5) are allowed in the SM, only 2$\nu$2EC has been experimentally observed until now. Geochemical studies exist for Ba-130 [38] and direct measurements of 2$\nu$2EC are reported for Kr-78 [39] and Xe-124 [40,41] with half-lives of the order of $10^{20}$–$10^{22}$ years. The continued searches for 2$\beta^+$, EC$\beta^+$ and 2EC processes are also well-motivated by the studies of nuclear structure models and advancements in experimental detection of their signatures. Experimental data from double beta decay experiments provide valuable constraints on theoretical models and calculations of nuclear matrix elements (NMEs), aiding in the validation, refinement, and deeper understanding of the underlying nuclear physics governing double beta decay processes. The hypothesised neutrinoless modes of these processes are expected to have half-lives significantly longer than the usual 0$\nu$2$\beta^-$ decay, with phase-space factors about 3–5 orders of magnitude smaller [33,42]. However, the rate of 0$\nu$2EC could be much bigger thanks to a resonant enhancement [33,43,44] and is therefore considered to have great potential for discovery of the Majorana mass of the neutrino. If observed, the neutrinoless positron emitting processes could help to distinguish models beyond the SM that can induce nonstandard contributions underlying the lepton number-violating signal [33,34,37].

## 3 Detector technology

Liquid scintillator detectors are a well-established technology in neutrino physics. They generally feature high light yield and therefore good energy resolution and a low energy threshold. Moreover, they reach high radiopurity, have low levels of optical impurities, lend themselves to isotope loading, and allow to cover large volumes at reasonable costs [45,46]. In recent years, novel ideas in the field of liquid scintillator technology emerged – most prominently hybrid and opaque scintillators – which now add unprecedented particle discrimination power at MeV-scale to the list of their benefits [47].

### 3.1 Hybrid scintillator

As illustrated in Fig. 1, hybrid scintillators exploit the particle-dependent ratio of Cherenkov and scintillation light (Č/S ratio) to discriminate between particles or event types [9]. More massive particles produce less Cherenkov light, because they are closer to the Cherenkov threshold [48,49]. Likewise,

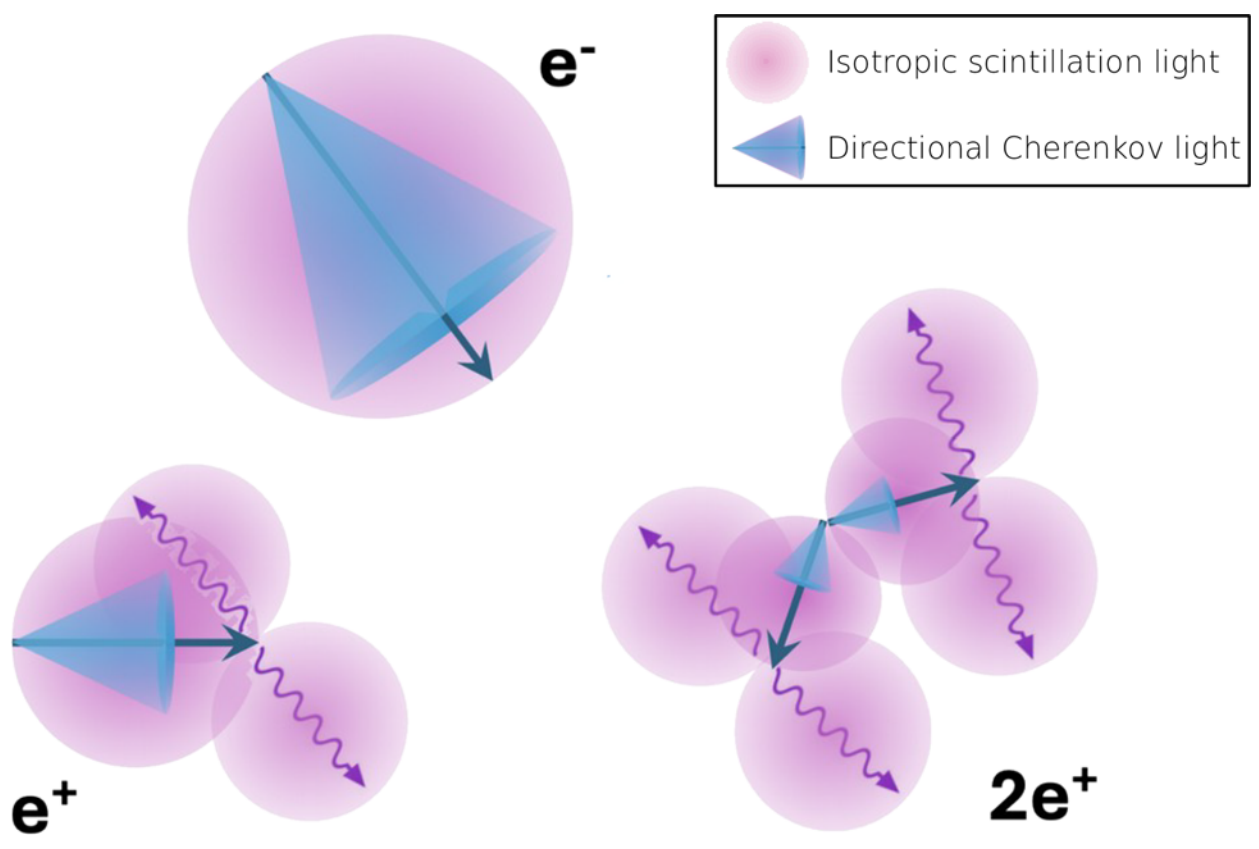

**Fig. 1** Illustration of the proposed strategy of particle discrimination in hybrid scintillator based on the ratio between the intensities of Cherenkov (blue) and scintillation light (violet)

gamma-rays produce only low amounts of Cherenkov light, as they transfer low amounts of energy into several recoil-electrons via multiple Compton-scatters. These Compton-electrons are often close or below the Cherenkov threshold. For positrons, the amount of Cherenkov light is lower with respect to an electron event, because the total amount of visible scintillation light $E_{\text{vis}}$ in a detector includes $2 \cdot 511$ keV energy from two annihilation gammas, which produce very little Cherenkov light, as explained above. Therefore, the kinetic energy of the positron is lower than the overall energy of the event. For instance, a positron of 2 MeV kinetic energy features the Cherenkov signal of an electron of same kinetic energy, but causes about 50% more scintillation light due to the additional two annihilation gammas of $2 \cdot 511$ keV.

Currently, two hybrid approaches exist: In the water-based hybrid approach, 1–10% scintillator is combined with water in small micelles via a surfactant interface [16,17,50]. This approach tunes down the amount of scintillation light, such that the fast Cherenkov-light can be separated from the scintillation pulse using fast electronics. While this approach yields a cost effective and safe detector medium with respect to flammability, it also diminishes the overall light yield due to the small amount of scintillator present in the medium. A newer approach is the hybrid-slow approach, which exploits intrinsically slow fluors or solvents [51–53]. Here, the full scintillation light is delayed in time such that the small peak of Cherenkov-light becomes visible during the beginning of the light emission. This approach offers the full scintillation light yield and therefore the good energy resolution and low energy threshold of classic liquid scintillators at the cost of spatial resolution. In NuDoubt$^{++}$, we introduce opacity to the slow hybrid scintillator, which offers high spatial resolution and introduces additional possibilities for particle identification.

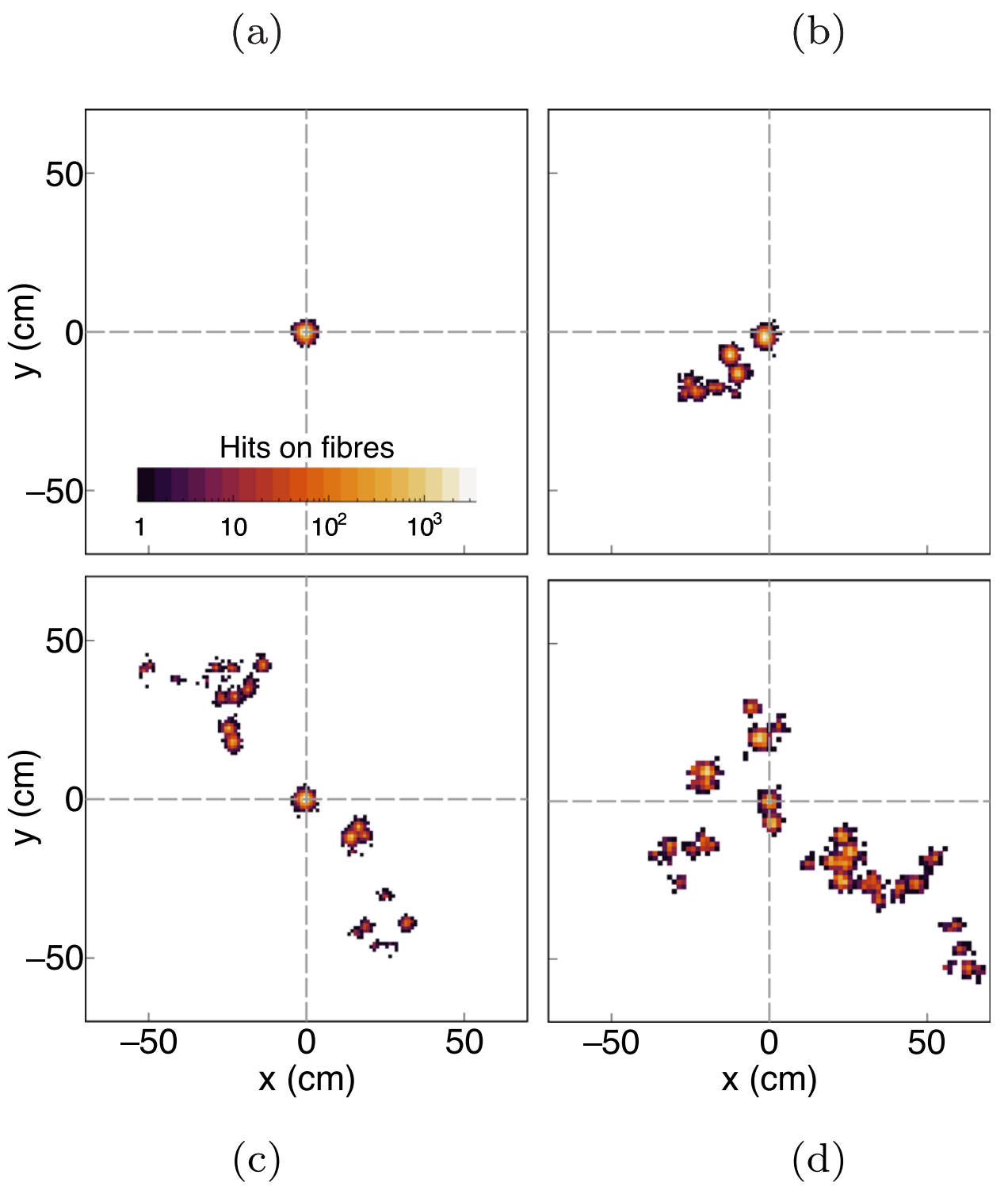

**Fig. 2** Illustration of the proposed strategy of particle discrimination in opaque scintillator based on the topology of hits of optical photons per fibre. Geant4 simulations of **a** an electron of 2.8 MeV kinetic energy, **b** a gamma-ray of 2.8 MeV energy, **c** a single positron of 1.78 MeV kinetic energy, and **d** two positrons of 0.38 MeV kinetic energy each. All four events deposit a total energy of 2.8 MeV in the detector. The simulation uses a scintillator composed of 98 wt.% solvent, and 2 wt.% wax, which has a 2 mm scattering length, a 2 m absorption length, and a light yield of 9000 photons/MeV. The fibre array uses a triangular pattern with 1 cm pitch. Fibres have a 3 mm diameter and are running parallel to the $z$-axis. All particles are generated at $(x, y, z) = (0, 0, 0)$

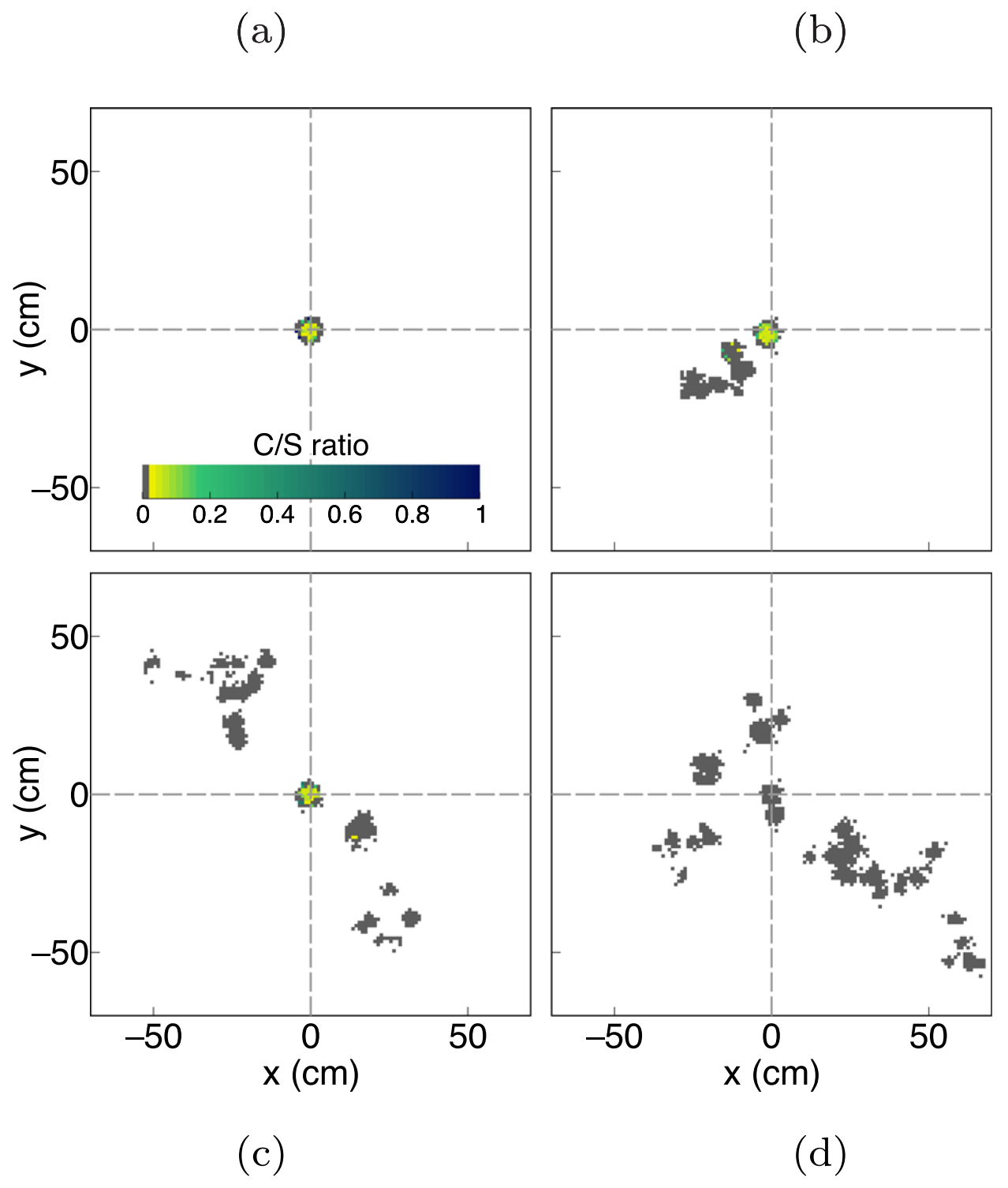

**Fig. 3** The ratio of hits of Cherenkov and scintillation photons for each fibre is displayed for the events in Fig. 2. Hit fibres with Č/S = 0 are represented in grey on the colour scale. Fibres without hits have white colour

### 3.2 Opaque scintillator

Opaque scintillators confine scintillation light through Mie scattering of optical photons [10,54]. The local light depositions are then picked up by a dense grid of wavelength-shifting fibres. The fibres are running through the entire detector volume and guide the light to both their ends, where it is read out with silicon photomultipliers (SiPMs). A opaque detector therefore retains the topological information, where ionising particles have deposited their energy, because the scintillation light stays close to the interaction points of the particles. Since each type of ionising particle leaves a different topological pattern of energy depositions in the scintillator medium, they can be distinguished. Electrons deposit their energy in a short ionisation trail and create a single blob (**b**ulky **l**ight **ob**ject) of scintillation light of a few centimetres size (Fig. 2a), gammas leave several blobs as they do multiple Compton-scatters followed by a final photoelectric effect (Fig. 2b), and positrons combine the patterns of electrons and gammas, because they first lose their kinetic energy in an ionisation trail like an electron and then produce two annihilation gammas, which leave their characteristic patterns (Fig. 2c).

Opacity in the scintillator medium can be achieved by blending classic transparent scintillator with wax, yielding a scintillator called NoWaSH [54]. Recently, it was possible to reduce the wax content of NoWaSH by an order of magnitude to just 2 wt.%. Due to this negligible fraction of non-scintillating wax, novel NoWaSH formulations reach similar light yields as classic transparent scintillators [55].

It is important to recognise that the energy of an annihilation gamma can be fully reconstructed even if it is not fully contained in the detector volume, e.g. when the positron annihilation happens close to the border of the detector vessel. It is merely required that the first two Compton-scatter vertices are fully contained and that their blobs can be separated from the ionisation blob of the positron and from each other. In this situation, one can reconstruct the full energy of the annihilation gamma by first measuring the scattering angle $\theta_\gamma$ at the first Compton-scatter blob, taking into account the positions of the positron blob and the second Compton-scatter blob. By measuring the energy $\Delta E_\gamma$ deposited in the first Compton-scatter blob, one can then determine the original energy $E_\gamma$ of the gamma exploiting the relation

$$\Delta E_\gamma = E_\gamma - \frac{1}{\frac{1}{E_\gamma} + \frac{1-\cos(\theta_\gamma)}{m_e c^2}} \tag{6}$$

where $m_e$ denotes the rest mass of an electron, and $c$ denotes the vacuum speed of light [56].

By the same principle, the trajectory of individual annihilation gammas, or at least their starting portions, can be reconstructed by relating the deposited energy in each Compton-scatter blob to the corresponding scattering angle and iteratively moving from blob to blob. With this approach, even the number of annihilation gammas in a double positron decay could be counted. This counting becomes particularly important, since positrons can form a bound state of positronium before annihilation. About half of the positronium is present as ortho-positronium with a lifetime of about 4 ns in liquid scintillator, which can decay into two or three gammas in matter [57,58].

Since annihilation gammas always carry 511 keV, even the measurement of neutrinoless decays at the endpoint of the decay energy spectrum is feasible although the annihilation gammas are not fully contained. In this situation, the measurement of kinetic energy of the positron deposited in its ionisation trail is sufficient.

### 3.3 Combining hybrid and opaque scintillators

The principle idea of our detector concept is the first-time combination of hybrid and opaque scintillator techniques in the same detector. This combination of media allows to combine the complementary background rejection capabilities of both detector concepts as discussed in Sects. 3.1 and 3.2. In addition, the combination offers novel insights to event topologies as shown in Fig. 3. This technique is based on the blob-dependent Č/S ratio instead of the particle-dependent ratio discussed in Sect. 3.1. Using this blob-dependent ratio, it becomes easily possible to determine, e.g. the primary vertex of a complex topology as shown in Fig. 3c or to tell several overlapping blobs of lower energy apart from a single blob of higher energy within a complex event topology. This technique is also aiding the reconstruction of events close to the boundary of an opaque detector, where incomplete topological information might be available.

The combination of hybrid and opaque media becomes possible through the recent progress of extending NoWaSH to cover a variety of solvents, including those of the hybrid slow solvent scintillator, through the use of novel wax types [55]. Since our detector medium is hybrid opaque NoWaSH, high levels of isotope loading become possible. Usually, such high isotope loading would degrade the transparency of a classic scintillator to a point where it becomes unusable [46]. However, in an opaque medium, the requirements on transparency are vastly relaxed due to local light read-out in combination with a scattering length of about 1 mm, which leads to much shorter average paths lengths of optical photons [10,54]. We describe our strategy for isotope loading in Sect. 4.

### 3.4 Expected performance

Despite our background discrimination strategy based on combined hybrid opaque scintillator, in the beyond standard model cases of $0\nu\mathrm{EC}\beta^+$ and $0\nu2\beta^+$, the most important background rejection strategy is the definition of a very small energy window, the region of interest (ROI), around the mono-energetic line from the positron(s). To achieve a small ROI, the energy resolution of the detector, which widens the mono-energetic lines of neutrinoless decay modes to peaks, has to be very good. Based on the preparatory work and the considerations made above, we propose a hybrid-slow opaque scintillator consisting of 88 wt.% linear alkylbenzene (LAB), 10 wt.% diisopropylnaphtalene (DIN), and 2 wt.% wax with diphenyloxazole (PPO) at a concentration of 1.0 g/l as wavelength shifter. This scintillator achieves a light yield of 9000 photons/MeV, an attenuation length of 17 m, and a scattering length of 2 mm [52–55]. Due to its slow scintillation emission with primary (secondary) decay-time constant of 13 ns (26 ns), the scintillation light is strongly delayed with respect to the Cherenkov peak. The scintillator achieves a purity of the Cherenkov light in the Cherenkov peak of 80% [52,53]. Current opaque detector readout technology is expected to reach between 360 and 550 PE/MeV (**P**hoton **E**quivalences) [25]. The largest loss of light happens at the interfaces between scintillator and wavelength-shifting fibres, as the geometrical capture efficiency via total internal reflection of the isotropically emitted shifted light is under 10%. We plan to improve this loss by using optimised wavelength-shifting light guides (OWL-fibres). These OWL-fibres feature an up to four times increase capture efficiency due to an optimised geometrical distribution of wavelength-shifter in the light guide [59] and are detailed in Sect. 5. Thus, a light yield of more than 800 PE/MeV appears possible in NuDoubt$^{++}$. This corresponds to an energy resolution better than 2% in the ROI.

## 4 Isotope candidates

In the search for beta plus decays, three isotopes are of particular interest: Kr-78, Cd-106, and Xe-124 (cf. Table 1). All have relatively high maximal detectable energies

$$E_{\max} :\approx Q_{2\beta^+} + 4m_e \approx Q_{\mathrm{EC}\beta^+} + 2m_e \approx \Delta M \tag{7}$$

in NuDoubt$^{++}$, where $Q_{2\beta^+}$ ($Q_{\mathrm{EC}\beta^+}$) denotes the atomic Q-value of the $2\beta^+$ (EC$\beta^+$) decay, $m_e$ the rest mass of an electron/positron, $\Delta M$ the mass difference between the neutral parent and daughter atoms, and where the atomic binding energy of electrons has been neglected. $E_{\mathrm{max}}$ is well above the gamma-line from Tl-208 at 2.614 MeV for all isotopes, leading to low background rates. As for the half-lives of these isotopes concerning neutrinoless and two-neutrino decay channels involving positron emission, only lower limits of $10^{20}$ to $10^{21}$ years have been established thus far. In the following, we discuss loading for the noble gases krypton and xenon in Sect. 4.1 and for the metal cadmium in Sect. 4.2.

### 4.1 Noble gases

Among the noble gases, krypton and xenon stand out as promising candidates for double beta decay studies. Krypton features two double beta isotopes and has been explored much less than xenon in this context. Kr-86 is a $2\beta^-$ isotope with low Q-value of 1.256 MeV and is henceforth neglected in this article [60]. Kr-78 decays to Se-78 via positive double weak interaction with a $E_{\mathrm{max}}$ of 2.881 MeV. The expected half-lives for the two-neutrino 2EC and EC$\beta^+$ decays in this isotope are about $10^{22}$ years, making the latter readily detectable in a small NuDoubt$^{++}$ detector. Two-neutrino $2\beta^+$ has an expected half-live of $10^{26}$ years, being reachable in an upgraded version of NuDoubt$^{++}$. The expected half-lives for the neutrinoless double beta decay modes start from about $10^{28}$ years for effective neutrino masses of few tenths of eV [35,37].

Similarly, xenon has two double beta isotopes. Xe-136 is a $2\beta^-$ isotope with a Q-value of 2.459 MeV and heavily explored by existing experiments [4,7]. Xe-124 decays to Te-124 via positive double weak interaction with a slightly lower $E_{\mathrm{max}}$ of 2.857 MeV. The $2\nu$2EC mode has been recently observed with a half-live of $1.1\cdot10^{22}$ years [40]. The expected half-lives for the two-neutrino 2EC and EC$\beta^+$ decays are in the range of $(0.4-97)\cdot10^{21}$ years, indicating that EC$\beta^+$ could also be detected in our experiment. Two-neutrino $2\beta^+$ has an expected half-live of $10^{27}$ years, being again reachable in an upgraded version of NuDoubt$^{++}$ [37]. The expected half-lives for the neutrinoless double beta decay modes start from $10^{27}$ years for effective neutrino masses of few tenths of eV [37,70].

While natural abundances of Kr-78 and Xe-124 are low, as gases they lend themselves to isotope enrichment in centrifuges and abundances of 99% are technically feasible. The solubility of noble gases in organic liquids is relatively high, especially for the heavier species [71]. It was shown that light yield and transparency of scintillator are only marginally affected by this amount of loading [72]. The dissolved concentration can be further enhanced with higher gas pressure, as described by Henry's law [73]. Measurements for organic solvents like toluene demonstrate that by raising the pressure to 5 bar, five times more krypton in the scintillator compared to normal pressure can be dissolved. While we expect the effect of higher loading on the optical properties of liquid scintillator to be minimal, it is necessary to demonstrate and quantify the effects. For this, we are currently constructing an overpressure test cell designed to determine the transparency, light yield, and krypton loading factor of our liquid scintillator as function of pressure [74]. The loading factor can be measured using the rate of low energy single beta minus decays of Kr-85. It has limited impact in our hybrid opaque detector and will later be greatly reduced by Kr-78 enrichment. To determine the absolute loading factor of krypton in our test cell, the fraction of Kr-85 in the krypton gas used for loading can be determined using a low-radioactivity proportional counter of calibrated volume [75,76].

### 4.2 Cadmium

The isotope Cd-106 is another suitable candidate for the search for positive double weak interactions. It decays into Pd-106 and has a high $E_{\mathrm{max}}$ of 2.775 MeV. The two-neutrino 2EC and EC$\beta^+$ double beta decays have expected half-lives of $(3-41)\cdot10^{20}$ years and $EC\beta^+$ is thus quickly detectable in our proposed experiment. Two-neutrino $2\beta^+$ decay has an expected half-life of $9.5\cdot10^{25}$ [37,77]. The half-lives for the neutrinoless double beta decay modes start from $10^{25}$ years [37,44].

While enrichment of up to 66% has been demonstrated in the search for double beta decays, the natural abundance of Cd-106 is small at 1.25% [68]. Regardless, for any actual abundance, a high loading factor is beneficial. Cadmium compounds are solid and therefore loading of the scintillator cannot be done via increased pressure as for noble gases. Traditional approaches load liquid scintillators with metals like cadmium by dissolving one of its compounds, typically an organometallic complex, which is often challenging due to lacking solubility [46]. Additionally, increased dissolution of metal complexes in liquid scintillators could compromise transparency, a crucial factor in classic large-scale scintillation detectors. Conversely, utilising opaque scintillators allows for simpler dispersion of the desired isotope within the scintillator, facilitating higher loading capacities and potentially enhancing opacity further. When using wax-based scintillators, which are manufactured as liquids and operated in a solid or highly viscous state [54], it is possible to disperse a cadmium compound throughout the scintillator rather than dissolving it. This technique allows to select from a larger variety of compounds, especially those which feature a higher fraction of cadmium. The chosen compound has to show a low absorbance either through high reflectivity or through high transparency throughout the UV and visible range of light. Highly reflective com-pounds aid the opacity of the medium and are thus preferable. For cadmium, it is even possible to use a scintillating compound which allows to increase the overall light yield of the scintillator. To demonstrate this loading technique, we disperse cadmium-tungstate ($CdWO_4$) powder (Sigma-Aldrich/Merck, -325 mesh) in a NoWaSH formulation made from 94 wt.% linear alkylbenzene (LAB), 1 wt.% of diphenyloxazole (PPO), and 5 wt.% of wax. This NoWaSH formulation has a typical light yield of about 9000 photons/MeV and its spectrum is dominated by the PPO emission spectrum peaking at 358 nm [54,55]. Cadmium-tungstate has a comparable typical light yield of 12,000 photons/MeV, shows a rather broad emission spectrum peaking at 490 nm and an absorption length of 0.6 m [68,78]. Cadmium-tungstate crystals have been successfully used in previous double beta decay searches [62].

**Table 1** Overview of the three $2\beta^+$ isotopes (first three rows) with maximal detectable energy $E_{\max}$ above the 2.614 MeV gamma-ray background of Tl-208 from the uranium/thorium decay chain. For comparison, three $2\beta^-$ isotopes with similar $E_{\max}$ are given (last three rows). Besides $E_{\max}$, measurements or limits on half-lives $T_{1/2}$ for the neutrinoless ($0\nu$) and two-neutrino ($2\nu$) decay modes from various experiments are listed [3–7,39–41,61–66]. For the limits, the respective confidence levels, C.L., are given. One can see that half-lives of beta plus decays are rather unknown in comparison to beta minus decays. In addition, the natural abundances $a_{nat}$ are listed. All isotopes can be acquired from several manufactures at enrichment fractions of more than 50%. For Ge-76, Cd-106, and Te-130, enrichment fractions of 88, 66, and 92%, have been achieved in the search for double beta decays, respectively [67–69]

| Isotope | $T_{1/2}(2\nu)$ /years | | | $T_{1/2}(0\nu)$ /years | | | C.L. /% | $E_{\max}$ /MeV | $a_{nat}$ /% |
|---|---|---|---|---|---|---|---|---|---|
| | $\beta\beta$ | $EC\beta^+$ | 2EC | $\beta\beta$ | $EC\beta^+$ | 2EC | | | |
| Kr-78 | $> 2.0\ 10^{21}$ | $> 1.1\ 10^{20}$ | $9.2\ 10^{21}$ | $> 2.0\ 10^{21}$ | $> 5.1\ 10^{21}$ | | 68 | 2.881 | 0.4 |
| Cd-106 | $> 1.7\ 10^{21}$ | $> 2.1\ 10^{21}$ | $> 3.1\ 10^{20}$ | $> 4.0\ 10^{21}$ | $> 1.2\ 10^{21}$ | $> 2.9\ 10^{21}$ | 90 | 2.775 | 1.3 |
| Xe-124 | | | $1.1\ 10^{22}$ | | | | | 2.857 | 0.1 |
| Ge-76 | $1.9\ 10^{21}$ | – | – | $> 1.8\ 10^{26}$ | – | – | 90 | 2.039 | 7.8 |
| Te-130 | $8.2\ 10^{20}$ | – | – | $> 2.2\ 10^{25}$ | – | – | 90 | 2.528 | 34.0 |
| Xe-136 | $2.3\ 10^{21}$ | – | – | $> 2.3\ 10^{26}$ | – | – | 90 | 2.459 | 8.9 |

Cadmium-tungstate is not hygroscopic, which eases its handling and combination with NoWaSH. We are able to disperse 50 wt% cadmium-tungstate powder (15 wt% cadmium) in NoWaSH. When combined, one can see the additional component from the emission spectrum of the cadmium-tungstate building up with increasing loading fraction using a fluorospectrometer. At the same time, the scattering length of the composite material becomes gradually shorter by a factor of two when going from 0 wt% to 50 wt% of cadmium-tungstate loading, aiding the opacity.

## 5 Optimised wavelength-shifting (OWL) fibres

For the light collection in the volume, we plan to use a new fibre concept optimised in terms of capture efficiency called Optimised Wavelength-shifting fibres (OWL-fibres) [79,80]. To maximise light capture, the wavelength-shifters (WLS) are located on the outer surface of the fibre since photons

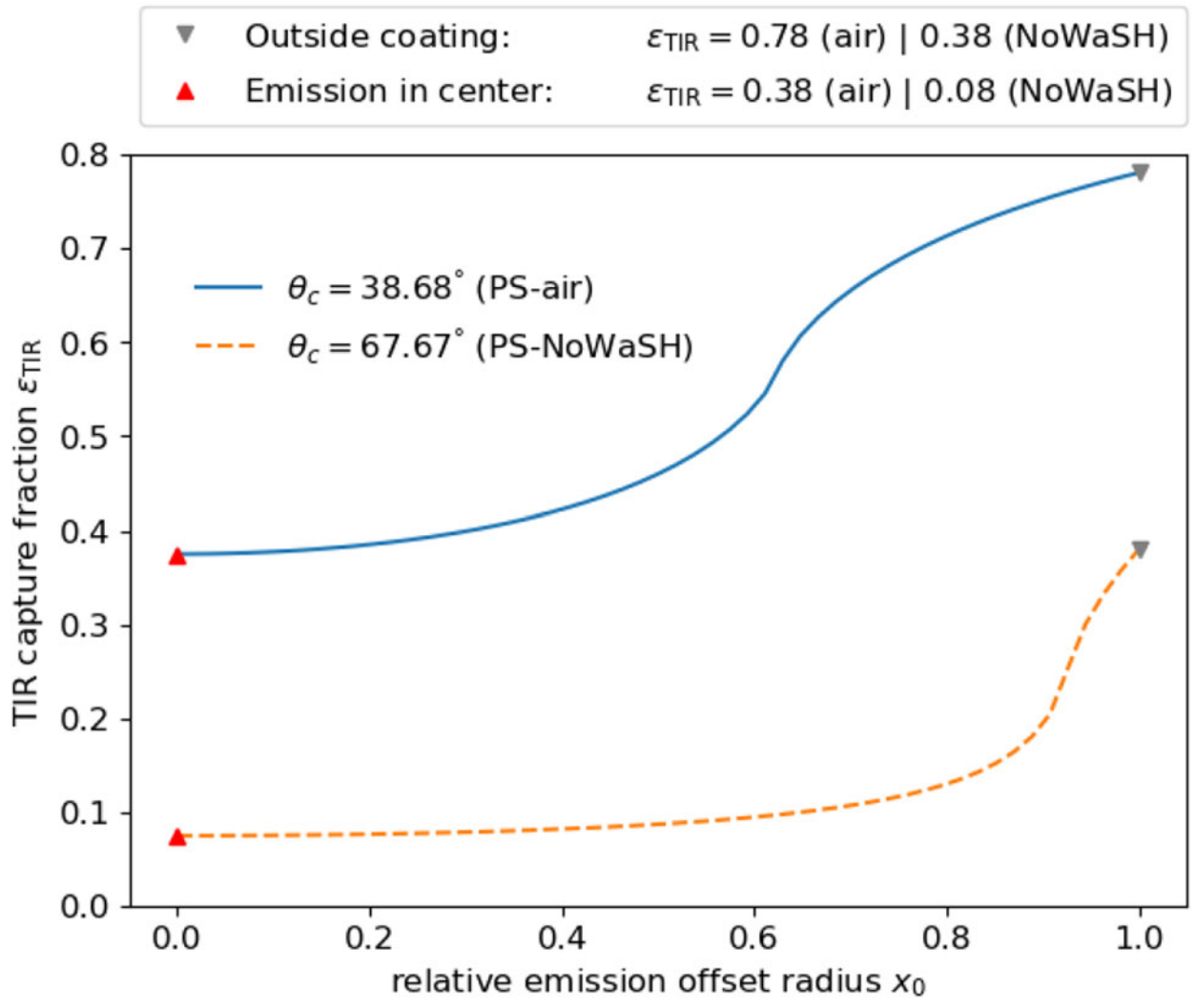

**Fig. 4** Fraction $\epsilon_{TIR}$ of the solid angle under which emitted photons are trapped in total internal reflection (TIR) as a function of the offset radius $x_0$ of the emission point. The two lines represent a polystyrene OWL-fibre immersed in air (solid line) or the scintillator NoWaSH (dashed line). Triangle markers indicate the minimum values for emission from the centre of the tube and the maximum values for emission from the outer edge of the tube

that are absorbed and emitted here have a higher chance of being captured by total internal reflection compared to those closer to the centre of the fibre [59]. The dependency of the trapping efficiency on the relative emission offset $x_0$ towards the surface is visualised in Fig. 4.

For polystyrene-based OWL-fibres in the proposed scintillator, the theoretical maximum trapping efficiency can reach up to 38%. This setup can trap nearly four times more photons compared to commercial available doped fibres where emission points are distributed throughout the fibre body. This will in turn lead to a substantial increase in light collection compared to current opaque scintillation detectors [25].

For the fabrication of the OWL-fibres, we strongly profit from expertise gained in the development of the Wavelength-shifting Optical Modules (WOMs) [81–88] over the last decade. PMT-coupled WOMs are developed for the detection of Cherenkov light in the IceCube Upgrade [87], while SiPM-coupled WOMs are used in the scintillation detectors of the SHiP experiment [85]. A WOM module is based on a optical quartz tube coated with wavelength-shifters (WLS) for the collection of light. Here, the WLS is applied on the outermost part of the quartz tube as a wavelength-shifting paint via dip coating, allowing homogeneous paint layers in the micrometre-range. The capture efficiency achieved with those tubes is at 80% of the theoretical maximum, and the attenuation length along the photon path reaches more than 3 m. The initial light loss of 20% is mostly caused by effects of self-absorption of the re-emitted photons in the paint layer [79].

We have performed first tests applying this technique to quartz rods with millimetre-range diameters. Compared to the larger-diameter WOMs, these showed a reduced attenuation length of around 2 m, likely caused by the decrease of the ratio of carrier material to the paint layer and its outer surface, which enhances effects like scattering and absorption in the paint layer and self-absorption by the WLS. For fibre lengths of 1–2 m as required for a 1-ton detector, the performance of the present OWL-fibres would still be competitive to commercially available fibres with attenuation lengths of 3–4 m due to the high fraction of initially captured light. Increasing the attenuation length of the first prototypes will further increase this benefit.

The first step to reduce the attenuation is an increase in the surface quality achieved by an advancement of the manufacturing process, which was originally designed for quartz tubes of about 10 cm diameter used for the WOMs. The upgrades include a better laminar flow control, to reduce surface contamination like dust in the coating station and a coating setup optimised for handling fibres with millimetre-range diameters to achieve a more homogeneous coating along the fibre.

Further to increasing the purity of the paint components which may improve scattering and absorption in the paint, we will study several ways to reduce the self-absorption in the paint layer: The use of special wavelength shifters with large Stokes' shift, e.g. DY-370XL with peak absorption (emission) at 360 nm (470 nm), would reduce the overlap between the WLS absorption and light emission spectrum [89]. Alternatively, quantum dots could be used instead of organic WLS. Quantum dots are crystals made of semiconductor material in the nanometre-regime. They show a narrow emission spectrum and a large Stokes shift between the absorption and emission spectra. This means that the overlap between the spectra is smaller compared to organic shifters reducing the losses due to self-absorption [90].

## 6 Backgrounds

The background for the EC$\beta^+$ ($2\beta^+$) searches arises from various processes capable of producing signals alike to one (two) positron(s) at energies approaching the $Q$-value of the isotope under investigation. The comprehension and control of possible sources of background is crucial for rare event search experiments like NuDoubt$^{++}$. In this section, we distinguish two different types of backgrounds.

*Internal background* events arise from the decay of radioactive isotopes within the detector volume itself, often stemming from impurities in materials such as the scintillator or the OWL-fibres. They are addressed in Sects. 6.1 and 6.2.

*External backgrounds* originate from any energy deposition occurring within or close to the detector, from particles originating outside this volume. Therefore, cosmic ray-induced backgrounds, radon gas-induced background, as well as natural radioactivity of the detector components (apart from scintillator and fibres), are considered external backgrounds. They are addressed in Sects. 6.3 and 6.4.

### 6.1 Internal scintillator background

Internal radioactive contaminations of the scintillator have the potential to impose relevant backgrounds to the measurement. However, organic liquid scintillators can be very effectively purified with standard chemical techniques like fractional distillation, a fact that has been demonstrated by both Borexino and KamLAND-ZEN [91,92]. For the materials in contact or close vicinity of the scintillator volume, screening and careful selection of manufactures allows to reach low background levels. The LAB-based liquid scintillators in the Daya Bay experiment achieved radiopurities for uranium (U) and thorium (Th) of $< 10^{-12}$ g/g and the JUNO experiment is expecting to achieve concentrations in the $10^{-15} - 10^{-17}$ g/g range [93,94]. Contaminations of potassium in wavelength shifters like PPO are reported below 0.5 ppm, while contaminations of uranium and thorium in PPO are below 1 ppb [95]. Those contamination levels can be further reduced in a purification step during scintillator production and lead to a contamination in the final scintillator of below ppb. Additional typical scintillator contaminations are C-14 and Bi-210. They are difficult to remove from the scintillator, but their decays are too low in energy and/or rate to be of relevance in our analysis, as seen in Fig. 5.

Another relevant source of background could stem from Rn-222 which is present as contamination in air or nitrogen gas cylinders.

Nitrogen is used as inert gas for scintillator vessels and during oxygen purging of scintillators to increase their light yield [45]. Measurements of the relative solubility coefficient of radon in nitrogen and different liquid scintillators show that about ten times more radon is present in liquid scintillators compared to nitrogen, when both are put in contact [96,97]. Typical nitrogen boil-off gas shows an activity from radon of about 50 μBq/m$^3$ which can be reduced through purification with activated charcoal by a factor of 100 [97]. This results in a typical radon activity in the scintillator of at most 5 μBq/m$^3$. Another source of radon is emanation from detector materials. This will be discussed in Sect. 6.4.

The wax used in the scintillator for opacity could be an additional source of background radiation. Wax samples where examined in a gamma spectrometer and no gamma-lines above the background of the spectrometer could be identified [54]. Upper limits on typical radio-impurities derived from this measurement are in the range of mBq/kg. Taking into account the loading fraction of wax in the scintillator of about 2 wt.%, rates much below mBq/m$^3$ are expected.

### 6.2 Internal background from OWL-fibres

A potentially very relevant contribution to the background inside the detection volume arises from the materials of the OWL-fibres interspaced in the detector volume. While the materials used in their manufacture are mostly hydrocarbons (PMMA, PEMA, Bis-MSB), they introduce traces of U-238, Th-232 and K-40 into the detection volume that cannot be shielded. Moreover, their surfaces will carry low amounts of dust particles that feature substantially higher U/Th concentrations. Table 2 denotes two contamination scenarios, baseline (100 ppt) and optimistic (1 ppt), that assume uniform contamination levels for all OWL-fibre materials (see below).

These internal background levels encompass both beta and gamma decays in the fibre materials. In the following, we differentiate between gammas emitted from the decays in the bulk volume of the fibres and beta decays mostly relevant when happening on or close to the surface of the fibres. We consider K-40 and the decay chains of U-238 and Th-232, assuming secular equilibrium for all isotopes and the usual branching ratios. We explicitly list the activity expected for the Tl-208 gamma line at 2.614 MeV and thus close to the endpoint of the $2\nu$ decay spectrum of Kr-78. In addition, we list a couple of high-energy $\beta^-$ emitters, Bi-212, Bi-214 and K-40, with spectra that overlap with the upper end of the Kr-78 spectrum. Note that the differing gamma and beta rates reflect the number of radioactive isotopes present in the relatively much larger bulk than surface volume of the OWL-fibres.

At the present state, no dedicated laboratory or manufacturing studies have been performed to evaluate the radiopurity levels that can be achieved in OWL-fibre production. Instead, we rely on prior experience from other low-background experiments, in particular JUNO and its pre-detector OSIRIS, to extrapolate the contamination and background levels that can be reached by careful material selection and manufacture in clean environments [98,99].

Preparatory screening studies performed for the components of the JUNO detector have shown that contamination levels substantially below 1 ppb weight can be reached in U-238, Th-232 and 0.1 ppb in K-40 for PMMA, bis-MSB, and other hydrocarbons based on careful selection of the manufacturing process [100]. Due to the chemical similarities, we assume that these levels are reachable as well for the PEMA used in the wavelength-shifting coating of the OWL-fibres. Note that radiopurity levels as low as 1 ppt have been reported for specially selected PMMA used in the NEXT and SNO experiments [101]. Indeed, both JUNO and its pre-detector OSIRIS specify uranium/thorium levels of 1 ppt or better [98,99] for the acrylics used in their scintillator vessels. Based on the span of contamination levels observed, we define here 100 ppt (10 ppt) as the *baseline scenario* and 1 ppt (0.1 ppt) as the most *optimistic scenario* for U/Th (K-40). The corresponding rates listed in Table 2 assume (conservatively) that all gammas emitted in these decays are detected within the active scintillator volume, while $\beta$ decays only contribute in the surface layer of the OWLs.

As lined out above, beta decays are as well induced by dust particles on the OWL-fibre surfaces. To avoid accumulation of dust on the OWLs, we assume that they will be rinsed with deionised water after production. Corresponding cleaning procedures developed in Borexino and JUNO show that – when performed in a sufficiently clean environment – the residual dust levels can reach levels corresponding to MIL-STD-1246 level 50 (e.g. [98,99]). The corresponding dust mass is 5 μg/m$^2$. Based on average levels of a US Geological Survey study, we assume 1.2 ppm of uranium, 7.2 ppm of thorium and 1.4 ppb of K-40 in the dust.

Using the dimensions of the OWL-fibres (length: 1 m, diameter: 3 mm, 10 mm pitch) and considering the amount of material from their bulk and in their matrix/cladding inside the detection volume, we can compute corresponding gamma rates from the OWL-fibre bulk volume and beta rates from their surfaces. Those rates are given in Table 2 for the baseline and optimistic scenarios.

### 6.3 Cosmogenic background

High-energy cosmic muons are a constant source of background in both surface and underground laboratories. In the present context, the relevant background arises in fact from secondary particle interactions, especially spallation on the nuclei of the detector materials (carbon) and surroundings.

The corresponding background signals typically exhibit temporal and spatial correlations to their parent muons.

Different strategies exist to mitigate these backgrounds. First and foremost, deployment of the detector within an underground laboratory is an efficient solution to reduce the cosmic muon flux [102]. In addition, the implementation of a dedicated muon veto system surrounding the target volume allows to tag residual muons with high efficiency, providing the possibility to use time and spatial coincidences to veto spallation backgrounds.

Muons can interact inside or prior to reaching the detector volume, either within the surrounding shielding materials, building structures, or in the rock of the laboratory, thereby generating fast neutrons [103]. These neutrons may penetrate the outer veto volume without notice and generate an uncorrelated background signal (proton recoil) in the detection volume. Fast neutron backgrounds can be mitigated based on passive shielding material, an extension of the active muon veto and/or particle ID. They are not considered to be a relevant background for this analysis.

The most recent results of KamLAND-ZEN800 have demonstrated that spallation isotopes from cosmic muons constitute an important background in the ROI that will become increasingly relevant for future high sensitivity searches [104]. The spallation nuclei are either created from carbon (in the scintillator itself) or the $\beta\beta$-isotope. Therefore, the second background component gains in prominence if loading factors are increased. Cosmogenic background can be reduced in several ways: positioning of the experimental setup in a deep underground lab like LNGS, SURF, SNOLab, or Jinping is the most efficient. In addition, short-lived isotopes (with half-lifed ranging from seconds to minutes) can be removed by forming a coincidence veto with the muon parent and other spallation products, especially neutrons. Borexino has demonstrated that such a three-fold-coincidence veto can be highly efficient [105]. This means that the shorter-lived carbon backgrounds can be suppressed by 1–2 orders of magnitude with an acceptable loss of experimental exposure. We apply a suppression factor of 20 for the short-lived C-10 ($\tau \sim 20\,\mathrm{s}$) and a factor 10 for the decays of the slightly longer lived C-11 nuclei ($\tau \sim 30\,\mathrm{min}$). Note that the situation is rather different for spallation products of the $\beta\beta$-isotopes [104]. Those can be much more long-lived and therefore no longer easily correlated to the parent muons. However, for most of the loading scenarios described in Sect. 8 they represent a negligible background compared to the C-10 and C-11 background rates.

### 6.4 Backgrounds from detector structures and laboratory

External gamma-rays from the surroundings of the detector and its outer layers impose a further background for rare-decay experiments like NuDoubt$^{++}$. Especially the high-energy gammas of Tl-208 and Bi-214 (daughter nuclei of U-238 and Th-232) can generate signal-like events around $E_{\mathrm{max}}$. During the detector design phase, material selection necessitates thorough radiopurity characterisation, with particular attention to materials in direct contact with the source, such as the vessels.

Rn-222 can emanate from the detector components or diffuse from the air of the laboratory. One of the primary concerns is the decay product Bi-214, which can produce two electrons by $(\alpha, \beta)$ disintegration with an energy around 3 MeV, and therefore constitute a source of background for double beta searches. This decay product is solid and can adhere to surfaces, including detector components. The target detector volume of the NuDoubt$^{++}$ experiment will be filled with liquid scintillator, which is an advantage compared with gaseous detectors regarding radon contamination. Nonetheless, a radon-free laboratory should be used during the construction phase, in order to avoid deposition on detector materials.

### 6.5 Background discrimination

Figure 5 illustrates the predicted event rates and spectra for relevant liquid scintillator backgrounds and the two-neutrino signals of krypton, assuming a detector placement at the Laboratori Nazionali del Gran Sasso (LNGS) (overburden of 3800 m water equivalent [102]). In this spectrum, event discrimination between electrons and gammas on the one hand and positron events on the other hand have been taken into account by applying reduction factors.

Opaque detector technology alone is expected to achieve background suppression factors of up to $10^3$ based on simple counting of blobs [10]. Through combination of hybrid and opaque detection techniques, we expect to significantly improve the suppression factors based on the clear separation of event categories via their Č/S ratio [106] as shown in Fig. 6 and based on the additional novel discrimination strategies discussed in Sect. 3.

In this article, we assume two scenarios of suppression power for the hybrid and opaque techniques as shown in Table 3. In the conservative scenario, we assume a suppression factor of just 100, which is applied to the decay rates of Po-210, C-14, Bi-210, Kr-85, electron recoils from solar B-8 neutrinos, and the gamma-ray background spectrum of the OWL-fibres. The two most important cosmogenic isotopes, C-10 and C-11, are $\beta^+$ emitters that cannot be discriminated effectively based on event topology or Č/S emission in the two-neutrino case. However, as mentioned in Sect. 6.3, the Borexino experiment (located at LNGS) has developed a three-fold coincidence veto based on parent muons and induced neutrons. From Ref. [105], we have extrapolated reduction factors of 90% and 95% for C-11 and C-10,

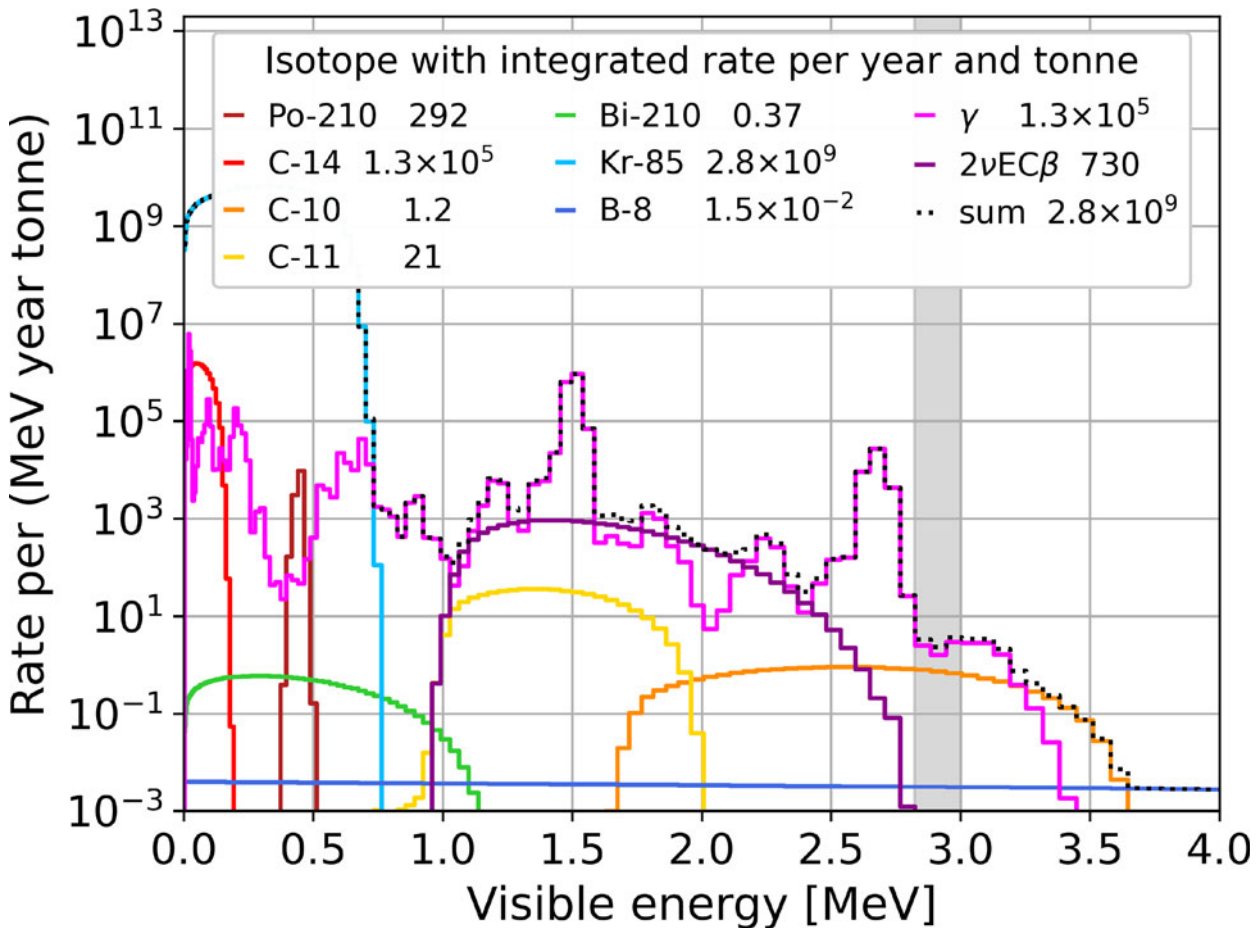

**Fig. 5** Expected spectra for a NuDoubt$^{++}$ detector located at the LNGS. Background and signal rates have been reduced by the conservative baseline suppression factors from Table 3. Their integrals are given in the legend. In addition, the expected signal spectrum for $2\nu EC\beta^+$ can be seen for a loading at 5 bar overpressure using 50% enriched Kr-78. Spectra are plotted with variable bin size reflecting the expected energy resolution. The grey band represents the region of interest (ROI) for the neutrinoless double beta decay of Kr-78 and extends from 2.82 to 3.00 MeV. The energy resolution in the ROI is 2.2%

**Table 2** Expected rates of relevant backgrounds from OWL-fibres in NuDoubt$^{++}$. The baseline (optimistic) scenario assumes 100 ppt (1 ppt) for uranium/thorium and 10 ppt (0.1 ppt) for K-40. All other isotope abundances assume secular equilibrium in the uranium/thorium decay chains. All rates are given in mBq per cubic metre of detection volume

| Location | Isotope | Rate /mBq/m$^3$ | |
|---|---|---|---|
| | | Baseline | Optimistic |
| OWL-fibre | U-238 | 103 | 1.03 |
| Bulk | Th-232 | 33.7 | 0.34 |
| ($\gamma$ only) | K-40 | 225 | 2.25 |
| | Tl-208 | 12.1 | 0.12 |
| OWL-fibre | Bi-214 | 2.1 | 0.028 |
| Surface+dust | Bi-212 | 0.48 | 0.014 |
| ($\beta$+$\gamma$) | K-40 | 0.5 | 0.005 |

respectively, ignoring for now a small accompanying loss in signal efficiency.

In the optimistic scenario, we study cut efficiencies for hybrid and opaque techniques based on Monte-Carlo simulations. Here, we apply a clustering algorithm derived from the Cambridge-Aachen jet clustering algorithm [107] to identify blobs and discriminate events based on their number of blobs. We further consider the Č/S ratio of event types by means of Gradient Boosted Decision Trees (GBDT) implemented from Scikit-Learn [108,109], taking into account the response of our slow scintillator and fibres in a detector simulation [106].

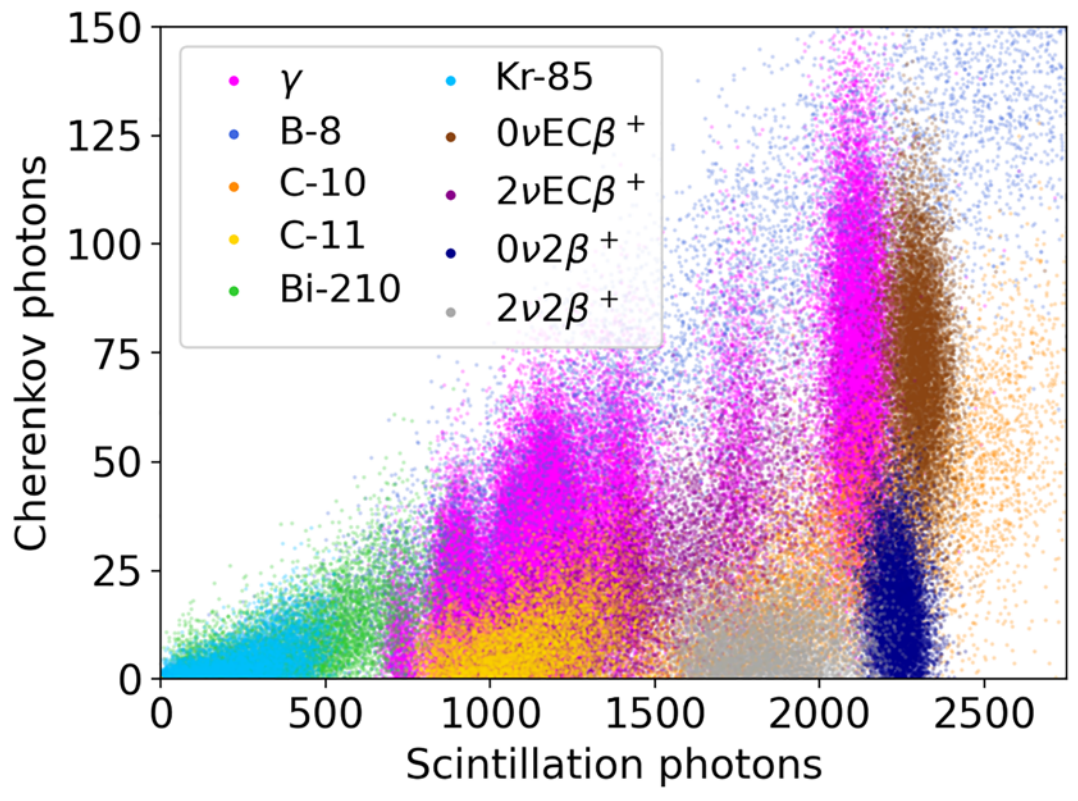

**Fig. 6** Simulation of the amount of Cherenkov photons and scintillation photons collected for several Kr-78 signal and background categories, assuming a light yield of 800 PE/MeV. A clear separation of the event categories can be seen. For better visualisation, 10,000 events (100,000 events) were simulated in each category except for gamma-rays (for gamma-rays). Actual event rates can be seen in Fig. 5

We calculate the expected energy resolution using a maximum likelihood approach to reconstruct energy from a full detector simulation. We focus on single blob events (electrons) originating from random positions within the detector. This inherently accounts for variations in the amount of collected light based on the emission point's distance to the fibre. Assuming a conservative light yield of 800 PE/MeV and 9000 photon/MeV, respectively, our results indicate that an energy resolution of approximately 2.2% can be achieved in the ROI.

The gamma-ray background above the 2.614 MeV gamma-rays from Tl-208, which fall inside the ROI for $0\nu2\beta^+$ and $0\nu EC\beta^+$ searches, stems from nuclear de-excitations of Po-214 after beta minus decays of Bi-214. With a rate of $1.34 \cdot 10^{-3}$ mBq/m$^3$ in the ROI after discrimination, these gamma-rays are the dominant background. It can be further reduced exploiting the novel discrimination technique based on the blob-dependent Č/S ratio introduced in Sect. 3.3. This discrimination has conservatively not been applied in this work, though.

## 7 Design of detector prototype

Based on the concepts discussed above, we consider here a prototype detector of about one metric tonne hybrid-slow opaque scintillator. This demonstrator can first be used to verify the hybrid and opaque particle discrimination capabilities of the new concept, as well as the complementarity of both discrimination strategies. For this purpose, the prototype will offer the possibility to deploy positron, gamma, and electron sources in its centre. We foresee a detector with cylindrical shape of equal diameter and height of roughly 110 cm. OWL-fibres will be running parallel to its symmetry axis, as illustrated in Fig. 7. While a spherical detector geometry offers the best volume to surface ratio, we propose a cylindrical geometry to ease the installation of OWL-fibres. Each OWL-fibre is connected with a SiPM on both ends to allow for a high light collection efficiency and therefore energy resolution, as well as positional resolution along the OWL-fibre exploiting signal delays between both ends. The arrangement of OWL-fibres just along one direction is sufficient for effective particle discrimination due to the complexity of the topological pattern of positrons compared to gammas or electrons, as shown in Fig. 2. The OWL-fibres are arranged in a triangular grid with a pitch of $\mathcal{O}(1\,\text{cm})$ between nearest neighbours, resulting in about 13,000 fibres in total. At this spacing, an average blob will illuminate about 15 OWL-fibres and yield an energy resolution per blob of 6%. From Monte Carlo simulations, we see that a cylindrical detector of one metric tonne hybrid-opaque scintillator is able to fully contain all electron events, but only 50% of positron and gamma events starting at its centre. Nevertheless, it is possible to reconstruct the full event of all positrons and gamma-rays even in this case because the energy of a 511 keV annihilation gamma-ray can be verified just from the angle of the first Compton-scatter and the energy deposited in the first Compton-scatter through Eq. (6), as explained in Sect. 3.2. The first two Compton-scatter vertices of annihilation gammas are contained for 97% of positrons in our detector, as we see in Monte Carlo simulations. In addition, the kinetic energy of positrons and electrons is always deposited in a single blob at the location of their creation inside the detector. The detector could be surrounded by an active veto, filled with transparent scintillator with good particle ID, to fully contain all signal events and especially to tag external backgrounds [110–112].

**Table 3** Cut efficiencies $\varepsilon_x = R_x^{\text{after}}/R_x^{\text{before}}$ for various signal and backgrounds types $x$, derived by comparing their rates before $R_x^{\text{before}}$ and after $R_x^{\text{after}}$ cuts. The baseline scenario assumes for most backgrounds a conservative overall suppression for the combined hybrid and opaque discrimination techniques of 100. An exception are the C-10 and C-11 backgrounds, where a threefold coincidence technique is uses, as explained in the text. In this article, the baseline scenario is used in our searches for two-neutrino decay modes. For neutrinoless decay modes, we use the optimistic scenario, where we evaluate hybrid and opaque discrimination performance in the ROI shown as grey band in Fig. 5. We evaluate the discrimination performance based on Monte-Carlo simulations for both neutrinoless decay modes separately. *For C-10 and C-11 backgrounds, the suppression factors from the threefold coincidence, as listed in the baseline scenario, apply additionally to the hybrid and opaque suppression factors in the optimistic scenario.
†Energies of the background do not extend into the ROI

| | | Cut efficiencies $\varepsilon_x$ | | | | |
|---|---|---|---|---|---|---|
| Scenario | | Baseline | Optimistic | | | |
| Signal type | | $2\nu2\beta^+$ and $2\nu\text{EC}\beta^+$ | $0\nu2\beta^+$ | | $0\nu\text{EC}\beta^+$ | |
| Discrimination technique | | All | Hybrid | Opaque | Hybrid | Opaque |
| Event type | Kr-85 | 0.010 | –† | –† | –† | –† |
| | Po-210 | 0.010 | –† | –† | –† | –† |
| | Bi-210 | 0.010 | –† | –† | –† | –† |
| | B-8 | 0.010 | 0.001 | $< 10^{-4}$ | 0.296 | 0.016 |
| | Gamma-rays | 0.010 | 0.030 | 0.016 | 0.676 | 0.477 |
| | C-10 | 0.050* | 0.278 | 0.237 | 0.439 | 0.823 |
| | C-11 | 0.100* | $< 10^{-4}$ | 0.237 | $< 10^{-4}$ | 0.823 |
| | C-14 | 0.010 | –† | –† | –† | –† |
| | $0\nu2\beta^+$ | 0.900 | 0.900 | 0.920 | 0.001 | 0.187 |
| | $2\nu2\beta^+$ | 0.900 | 0.900 | 0.920 | $< 10^{-4}$ | 0.187 |
| | $0\nu\text{EC}\beta^+$ | 0.900 | 0.001 | 0.237 | 0.900 | 0.823 |
| | $2\nu\text{EC}\beta^+$ | 0.900 | 0.001 | 0.237 | 0.900 | 0.823 |

After the verification step, the prototype could be isotope-loaded and deployed at an underground facility. As discussed in Sect. 6.3, the most relevant background for this search are spallation products of carbon and cosmic muons. In the following, we assume the deployment of the detector at Gran Sasso laboratory and use the corresponding muon, neutron, and isotope rates. For the loading, a small vessel of about 10 l fiducial volume (roughly 10 kg fiducial mass) will be inserted in the centre of the detector, holding loaded hybrid-opaque scintillator. The overall amount of Kr-78 isotopes can be increased by high pressure loading and enrichment as detailed in Sect. 4.1. The usage of a small high-pressure loaded central volume, instead of a larger loaded volume at normal pressure, allows to suppress backgrounds from fibres outside of the loaded volume due to their vertex. Pressure

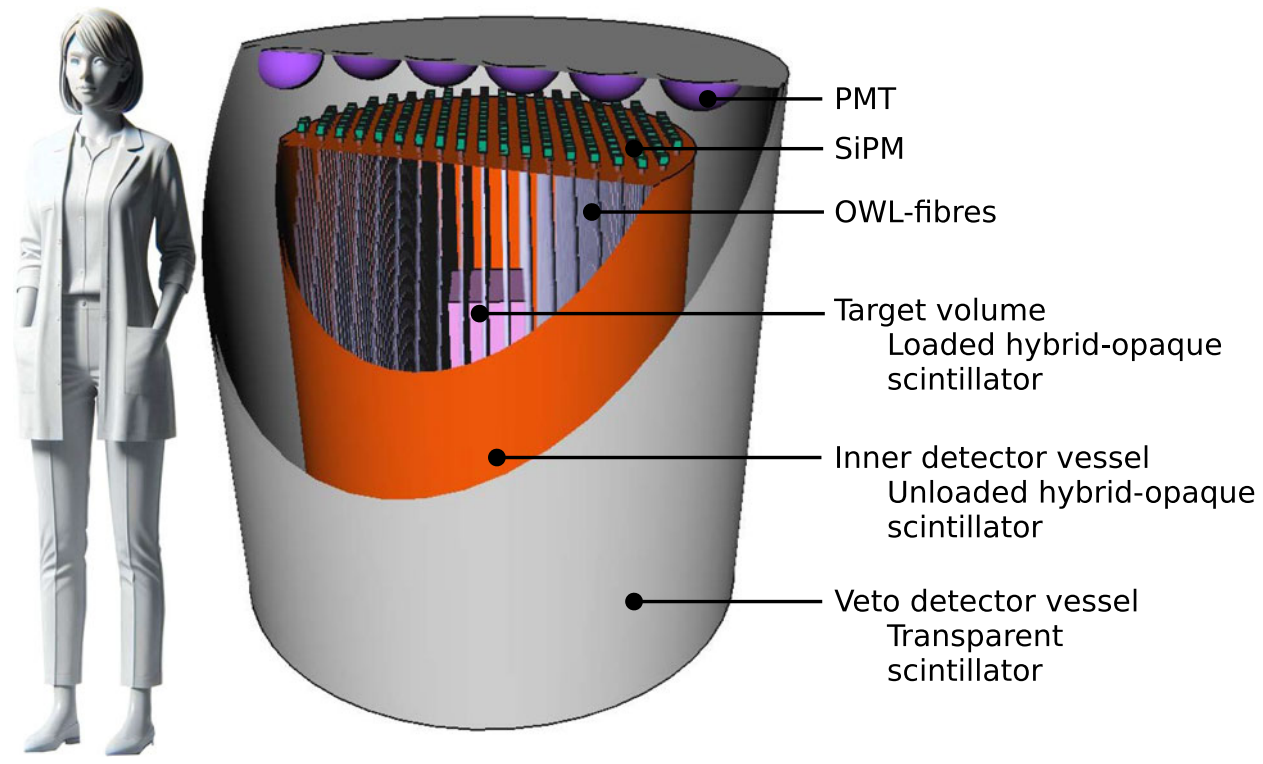

**Fig. 7** Basic detector design. From the outer to the inner layer: the active veto volume (grey) has a diameter and height of 155 cm, is filled with transparent scintillator and has PMTs (purple) on the top and bottom; the inner detector volume (orange) has a diameter and height of roughly 110 cm, holds 1 t of unloaded hybrid-opaque scintillator and parallel-running OWL-fibres (light grey) with a diameter of 3 mm and a fibre pitch of 1 cm (fibre pitch has been increased in the drawing for a better visualisation); dedicated SiPMs and electronics (green) are at both ends of each fibre. The inner cubical target volume (pink) is filled with about 10 kg loaded hybrid-opaque scintillator and has an edge length of roughly 10 cm

in the outer volume can be adapted to the pressure in the fiducial volume such that only the outermost vessel needs to be build from pressure-resistant materials like steel, which potentially contain a higher amount of radioactive isotopes. Due to the optical self-separation of the opaque NoWaSH scintillator medium, the outermost layers of OWL-fibres can be used as a self-veto against external backgrounds. Optionally and in addition, the fiducial and overall detector mass could be increased by elongating the cylinder along the symmetry axis. This way, a similar number of channels and, therefore, electronics and data acquisition system can be re-used to instrument the larger detector, while the energy resolution will decrease due to photon absorption in the fibres.

In the configuration with 10 l fiducial volume, the aim of NuDoubt$^{++}$ is the discovery of the standard model decay modes $2\nu EC\beta^+$ and $2\nu 2\beta^+$ as well as improvements of limits on beyond-standard-model decay modes $0\nu EC\beta^+$ and $0\nu 2\beta^+$. Due to the large relevance to the overall background budget, a radio-clean version of OWL-fibres based on a substrate selected for low radioactivity has to be used. In terms of isotope loading, three sequential phases can be implemented, (cf. Table 1) starting with high pressure loading of enriched krypton gas, then going to high pressure loading of enriched xenon gas (cf. Sect. 4.1), finally succeeded by enriched cadmium loading as described in Sect. 4.2. While the noble gases would be located in the central 10 l fiducial volume, for the solid cadmium-tungstate two loading options exist. It can either be dispersed throughout the same central 10 l volume as the nobel gases, or it could be placed at several confined locations between fibres, where it can be trapped using one of the high viscosity formulations of NoWaSH [55]. The latter option benefits from additional discrimination power based on vertices. In Sect. 8, we detail the physics reach of our detector configuration using krypton as an example.

Informed by the results from our first detector described above, the final step will be dedicated to beyond-standard-model searches of $0\nu EC\beta^+$ and $0\nu 2\beta^+$ at higher sensitivity. In this stage, an upgraded pressure vessel is foreseen to allow very high pressure loading while the detector volume and number of channels is not expected to increase significantly.

## 8 Sensitivity

We carry out two sensitivity studies for NuDoubt$^{++}$ using krypton as example and assuming an underground deployment at the LNGS with corresponding background rates as determined by the Borexino experiment [113]. The shape of the spectra is generated using the Geant4 BxDecay0 C++ library [114–117]. This is also used to generate the two-neutrino spectra shapes. For the gamma spectrum, the rates discussed in Sect. 6 are used. Energy values of gamma peaks were taken from the live chart of nuclides [118].

In our first study, we investigate the discovery potential towards two-neutrino double decay modes. In this study, we use the conservative baseline discrimination factors from Table 3 for all backgrounds. For the background rates from OWL-fibres, we assume the conservative baseline scenario from Table 2. We apply an energy dependent smearing for all spectra in accordance with our expected energy resolution. We apply this smearing through a Gaussian function, resulting in the spectra depicted in Fig. 5, with their binning reflecting the energy resolution. The expected sensitivity for $2\nu EC\beta^+$ is shown in Fig. 8. Here, the detector volume is set to one tonne, while the loaded central fiducial volume are 10 ls. For the signal rate not only the detector volume has to be considered, but also the level of enrichment and the pressure. The sensitivity study was carried out by calculating the minimum negative log-likelihood (LLH) based on a Poisson distribution in each bin [74]. The sum of all spectra (background and signal) is fitted under the assumption of the background-only hypothesis. This was done for different signal rates to give the typical LLH profile. Given more statistics, this value changes with the duration of the experiment as well as with the size of the detector. The minimal double decay rate that can be reliably identified in the data is obtained by examining the five standard deviation interval of the LLH. This rate can then be converted into a half-life, for which the signal can be detected at five standard deviation confidence level. In Fig. 8, one can see the relation between the number of Kr-78 atoms and the half-life. The shorter the half-life of the atoms, i.e. the higher their decay rate, the fewer atoms are needed to reach a given half-life. The verti-cal lines in the plot represent different loading scenarios, with each colour representing different levels of enrichment, with loading at 1 bar and with 5 bar overpressure. The horizontal dashed line represents the expected half-life of Kr-78. With a detector loaded with 50% enriched krypton at 5 bar overpressure in a 10 kg (scintillator mass) central fiducial vessel, NuDoubt$^{++}$ is able to reach this value with an exposure of less than one tonne-week corresponding to less than 20 kg-years. We find similar sensitivity for the $2\nu 2\beta^+$ decay.

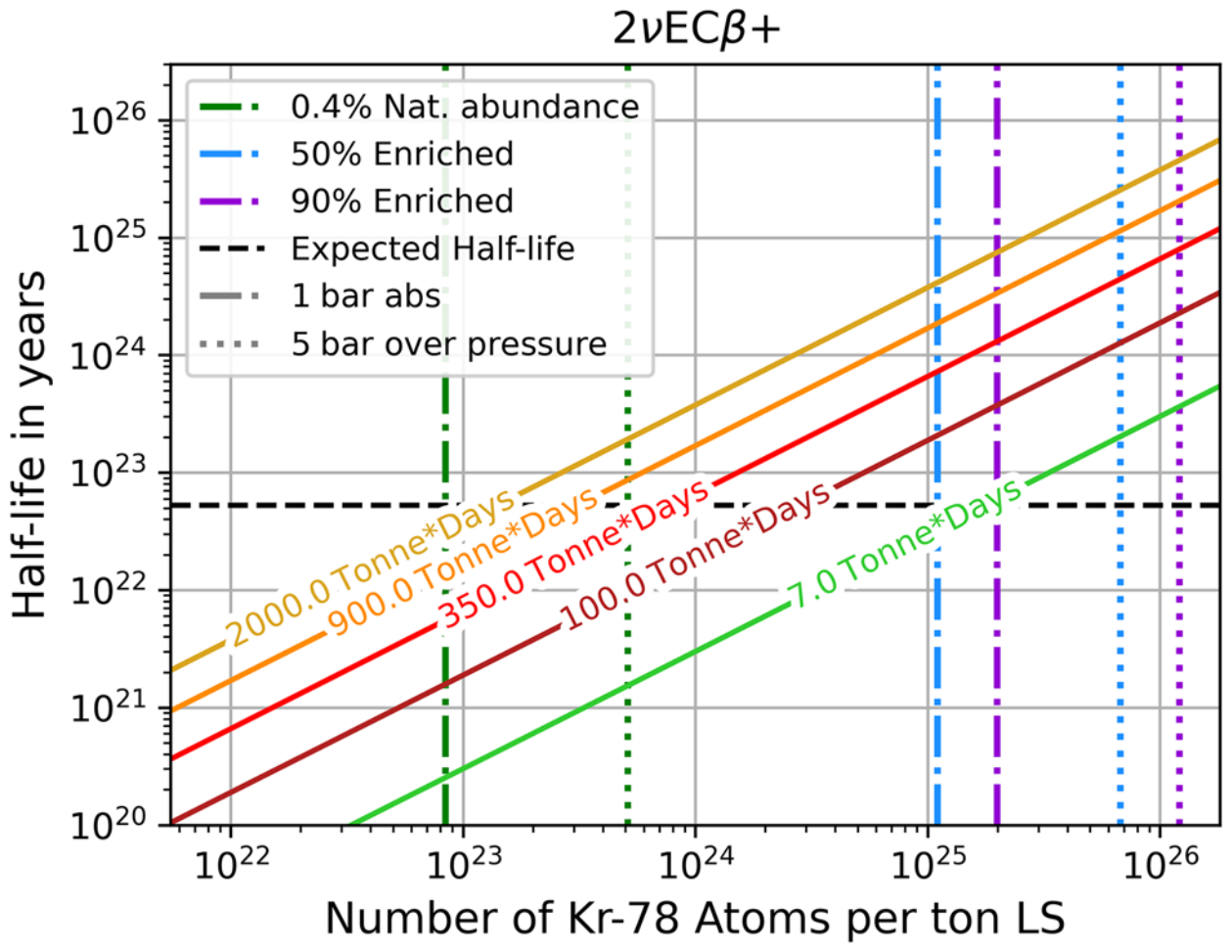

**Fig. 8** Discovery potential of NuDoubt$^{++}$ at five standard deviations significance for $2\nu EC\beta^+$. Similar values are achieved for $2\nu 2\beta^+$. Solid lines indicate the detectable half-life, depending on the minimum rate required to exclude the background-only hypothesis, for various amounts of decaying Kr-78 atoms per tonne of loaded scintillator mass. The minimum rate depends on the scintillator mass and the run time. Each solid line represents this relationship for a specific combination of tonne-days. The horizontal dashed black line shows the theoretical half-life of the $2\nu EC\beta^+$ decay [35], dashed vertical lines represent different loading and enrichment cases

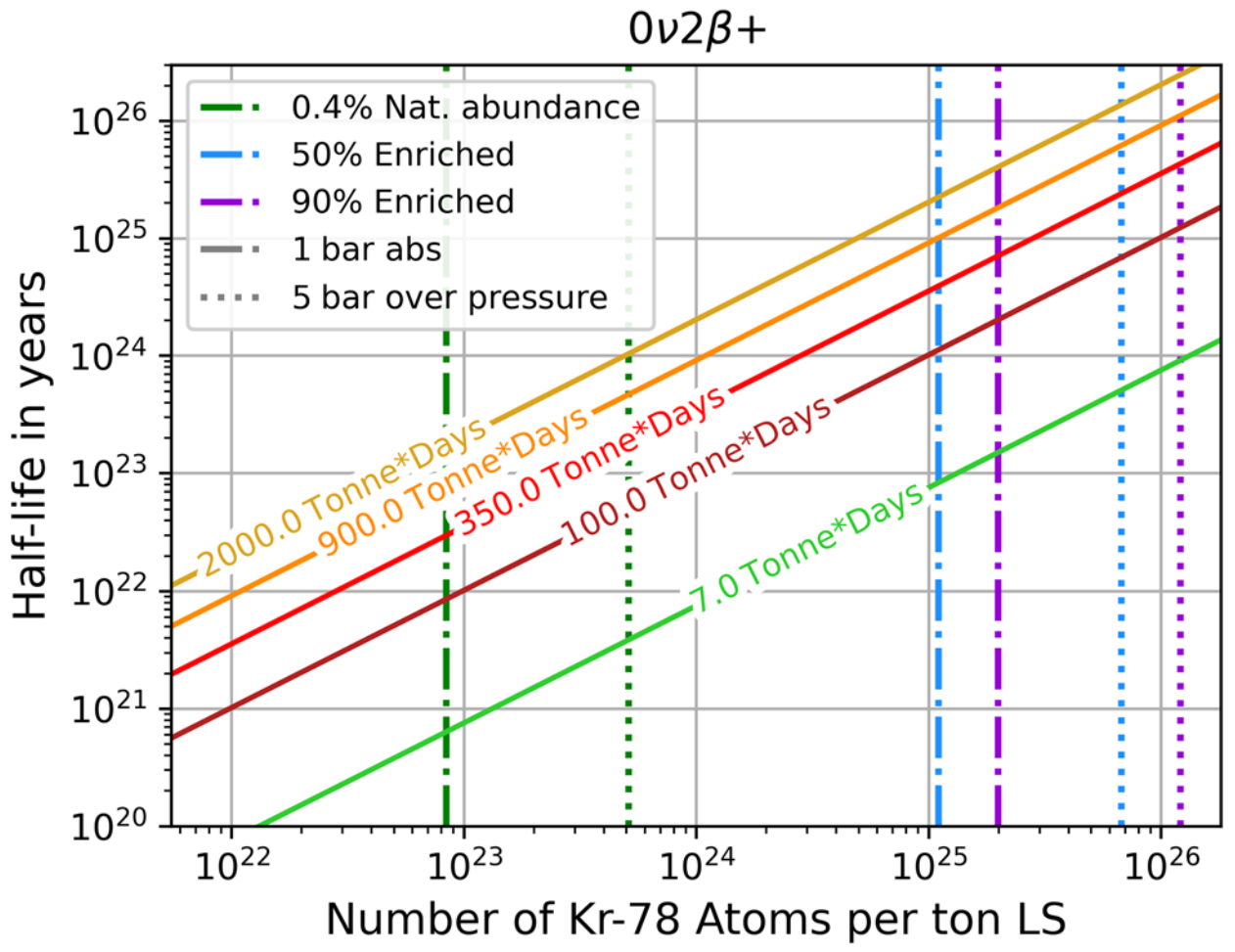

**Fig. 9** Exclusion sensitivity of NuDoubt$^{++}$ at 90% C.L. for $0\nu 2\beta^+$. Similar values are achieved for $0\nu EC\beta^+$. Solid lines indicate the longest rejected half-life, depending on the highest rate required to exclude the corresponding signal+background hypothesis. Half-lives below the solid lines are rejected. See caption of Fig. 8 for further explanations

Similar to the sensitivity study for the two-neutrino decay, a lower limit can be derived for the neutrinoless case. To do this, we only consider signal events around 2.881 MeV and compare them to the background in the ROI, depicted as grey band in Fig. 5. We define the ROI between 2.82 and 3.00 MeV, considering our energy resolution of 2.2% in the ROI. For this second study, we use the optimistic discrimination factors from Table 3 and the optimistic background rates for our OWL-fibres from Table 2. As seen in Fig. 9, with a detector loaded with 50% enriched krypton at 5 bar overpressure in a 10 kg (scintillator mass) central fiducial vessel, we can set a limit on $0\nu 2\beta^+$ of almost $10^{24}$ years at 90% C.L. after an exposure of one tonne-week or equivalently 20 kg-years. The background index inside the ROI, not considering signal efficiencies, lies at $2 \cdot 10^{-5}$ counts/(keV kg year) with respect to scintillator mass and $2 \cdot 10^{-3}$ counts/(keV kg year) with respect to Kr-78 mass, respectively. We can further improve the background index to $6 \cdot 10^{-9}$ counts/(keV kg year) with respect to scintillator mass and $7 \cdot 10^{-7}$ counts/(keV kg year) with respect to Kr-78 mass, respectively, by applying stricter cuts at the expense of overall signal rate. We can achieve a similar limit for the $2\nu EC\beta^+$ decay mode.

## 9 Conclusion

Exploring modes of double beta decay involving positron emission offers new opportunity for understanding nuclear structure. The observation of an additional double beta process, specifically the standard-model two-neutrino mode, would provide a data point for refining the calculations of nuclear matrix elements relevant to nuclear structure models. This observation could serve as an additional cross-validation of current computational methodologies. Additionally, the hypothesised neutrinoless modes of these processes hold promise for revealing nonstandard contributions to lepton number-violating signals and probing the Majorana mass of neutrinos.

Despite experimental challenges and the scarcity of suitable candidate nuclei, ongoing efforts to detect these modes are driven by their potential implications for particle physics. The rarity of these processes highlights the necessity for enhancing experimental sensitivity techniques. The novel approach combining advanced hybrid and opaque scintillator techniques presented in this work offers a potential breakthrough in search for double beta plus decays.

Hybrid scintillators utilise the ratio of Cherenkov and scintillation light to discriminate between particle types, while opaque scintillators confine light through multiple scattering of optical photons, allowing for retention of topological information about energy depositions from ionising particles. Additionally, opaque scintillators enhance traditional benefits of liquid scintillator detectors such as high light yield, excellent energy resolution, and low energy thresholds. Combining the background rejection capabilities of hybrid and opaque scintillator techniques in the same detector represents an advancement in double beta decay searches. However, achieving a small ROI for double beta decay modes beyond the standard model remains crucial, necessitating optimisation of energy resolution and overcoming losses at interfaces between scintillator and wavelength-shifting fibres. Optimised wavelength-shifting light guides, derived from novel wavelength-shifting optical modules developed within the IceCube Upgrade, offer potential for reducing these losses and enhancing our detector performance.

In a sensitivity study under conservative assumptions regarding background contamination, background discrimination and energy resolution, we see that already a moderately large detector of 10 kg krypton-loaded scintillator and an overall scintillator mass of one tonne can discover two-neutrino beta plus decay modes after two years of runtime. The same detector is also capable to improve current limits on neutrinoless decay modes by several orders of magnitude. Using a larger detector with tonne-scale fiducial volume, the expected half-lives of neutrinoless decay modes come into reach.

**Acknowledgements** Following the topical order of the article, we thank Lukáš Gráf for discussions on phenomenological aspects of double beta plus searches. We especially thank Anatael Cabrera for discussion on initial ideas regarding double beta plus measurements with the LiquidO technology. We further thank Hardy Simgen and Jonas Westermann for discussions and support regarding low-radioactivity proportional counters, Kai Loo and Arshak Jafar for providing simulations used for early studies on the Č/S ratio, as well as the authors of reference [52] (H. Steiger et al.) for granting us early access. We thank members of the IceCube, LiquidO, and Theia collaborations for feedback and discussion on various aspects of our concept.

**Funding** This work has been supported by the Cluster of Excellence "Precision Physics, Fundamental Interactions, and Structure of Matter" (PRISMA$^+$ EXC 2118/1) funded by the German Research Foundation (DFG) within the German Excellence Strategy (Project ID 390831469). We are especially thankful for the support of the PRISMA Detector Laboratory.

**Data Availability Statement** Data will be made available on reasonable request.

**Code Availability Statement** Code/software will be made available on reasonable request.